\magnification=\magstep1
\baselineskip=15pt
\vsize = 24.5 true cm
\def\refindent{\par\penalty-100\noindent\parskip=4pt plus1pt
                \hangindent=3pc\hangafter=1\null}

\centerline{\bf A SPECTROSCOPIC SURVEY FOR BINARY STARS }
\centerline{\bf IN THE GLOBULAR CLUSTER NGC~5053}
\bigskip
\bigskip
\centerline{Lin Yan\footnote{$^1$}{Present address: European Southern
Observatory; Karl-Schwarzschild-Str. 2; D-85748 Garching b. M\"{u}nchen;
Germany} and J. G. Cohen}
\centerline{Department of Astronomy, MS 105-24}
\centerline{California Institute of Technology; Pasadena, CA 91125}
\centerline{Email: lyan@eso.org, jlc@astro.caltech.edu}
\vskip 3.0 cm
\centerline {To appear in the October 1996 issue of AJ}
\vfill\eject

\centerline{\bf ABSTRACT}
\bigskip

We carried out a radial velocity survey for spectroscopic binaries 
in the low density globular cluster NGC~5053. Our sample contains a total of
77 cluster member giant and subgiant stars 
with visual magnitudes  of 
14.5$-$18.6. Of these 77 stars, 66 stars have on average of 3$-$4 
measurements with a total of 236 velocities. A typical velocity 
error per measurement
is $\sim$3~$\hbox{km s}^{-1}$.
The stars in our sample are spatially distributed from the 
cluster center out to 10 arcminutes in radius (4.5 core radii). 
Among these 66 stars with multiple velocity measurements, we discovered 
6 spectroscopic binary candidates.
Of these six candidates, one was discovered
as a binary previously by Pryor et al. (1991) and
candidate ST is a binary with a very short-period of three to
five days. We obtained three possible orbital solutions for binary
candidate ST by fitting its
radial velocity data. These orbital solutions are 
consistent with star ST being a cluster member,
although its spectrum has much stronger MgI triplet absorption lines than that
of a typical low-metallicity giant star. 

Using a Monte-Carlo simulation method, we estimated 
the fraction of binary systems which may have been missed from
our detection due to unfavorable orbital configurations. With our survey, 
the binary discovery efficiency is 29\%\ for systems
with 3~d $\le$ P $\le$ 10~yr, 0.125 $\le$ q $\le$ 1.75 and eccentric 
orbits (0 $\le$ e $\le$ 1). 
This yields a binary frequency of 29\%. 
We also applied Kolmogorov-Smirnov (K-S) tests to the cumulative 
distributions of maximum velocity variations from the actual
measurements and the synthetic velocity data. 
The results from these tests are consistent with 21$-$29\%\ 
binary population with 
3~d $\le$ P $\le$ 10~yr, 0.125 $\le$ q $\le$ 1.75 in NGC~5053.
The hypothesis of a binary frequency in NGC~5053 
higher than 50\%\ is rejected with a confidence level higher than 85\%.

The binary frequency in NGC~5053 derived 
from our survey is somewhat higher than estimates for other
clusters by various surveys. This is perhaps related to the fact that
NGC~5053 is relatively dynamically young compared to other clusters.
We also argue that the binary population in globular clusters is not
significantly deficient compared to binaries in other stellar environments 
such as open clusters, or to 
field and low metallicity halo stars.

\bigskip
\bigskip
\centerline{\bf 1. INTRODUCTION}
\bigskip

Recently it has been realized that a primordial binary frequency as small
as 3\% can fundamentally change the dynamical evolution of an entire
globular cluster (Heggie \&\ Aarseth 1992). 
Mass segregation caused by two body relaxation in a cluster
essentially transfers ``heat'' (stellar kinetic energy) from the
cluster core to the ``cooling'' edges, while simultaneously pushing
the cluster towards higher central concentration ({\it gravothermal
collapse}). The process of gravothermal collapse can be greatly
modified by a binary star population (Gao {\it et al.} 1991; Heggie
\&\ Aarseth 1992; McMillan \&\ Hut 1994; Vesperini \&\ Chernoff
1994). Gravitational binding energy in binary stars can be extracted
and converted into kinetic energy during encounters with other
stars. The extracted energy can supply a central heat source to
stave off or reverse gravothermal collapse. 

There is considerable observational evidence that
binary stars do exist in globular clusters 
(see the detailed review by Hut {\it et al.} 1992).
Discoveries of low-mass X-ray binaries (LMXB) in globular clusters (Grindlay
{\it et al.} 1984) first hinted that a primordial binary population
could exist in globular clusters, although the formation of LMXBs doesn't
necessarily require the pre-existing binary populations.  
Indisputable evidence came from the discovery of long-period binary 
millisecond pulsars
in low density globular clusters during the late 80's and the early 90's. 
Among a total of 
$\sim$40 millisecond 
pulsars in globular clusters, $\sim$12 of them are actually in binary systems. Furthermore,
two of these binary millisecond pulsars ---discovered 
in the {\it low density clusters} M~4 
and M~53--- have 
{\it periods as long as 200$-$300~days} (McKenna {\it et al.} 1988; 
Kulkarni {\it et al.} 1991).
Binaries with such long periods cannot be formed through single 
star-star tidal capture. In fact, PSR~1640-26 in M~4 is a triple system.
These two long period millisecond pulsars either
have to be formed by star-binary and binary-binary encounters, 
or their progenitors are primordial
binary stars. 

Recent direct searches for spectroscopic and short-period 
eclipsing binaries have
discovered many binary systems and binary candidates in globular 
clusters (Pryor {\it et al.} 1992;
C$\rm \hat o$te {\it et al.} 1994; Mateo {\it et al.} 1990; Yan \&\ Mateo 1994; 
Yan \&\ Reid 1995).
One of the drivers behind these efforts is to 
understand many apparently
different phenomena in globular clusters which may be intimately 
related to binary star population.
For example, such phenomena include blue stragglers,
color/stellar gradient
in the cores of post-core-collapsed clusters (Djorgovski \&\ King 1986), 
X-ray sources and
numerous ``recycled'' radio pulsars (Phinney 1992;
Phinney 1995).
The discovery of short-period eclipsing binaries among both blue
straggler stars and main-sequence stars (Mateo {\it et al.} 1990; 
Kaluzny and Krzeminski 1993; 
Yan \& Mateo 1994; Yan \&\ Reid 1995) suggests
that the short-period binaries may be
the progenitors of blue stragglers. 
Mass transfer/merger among stars in binary systems and stellar encounters
involving binary stars
are important processes for the
formation of blue stragglers. Also, numerous
X-ray sources and radio pulsars in globular clusters can be 
easily explained in terms of
binary systems containing degenerate stars. 

While mounting evidence, direct or indirect, has suggested
the existence of both short and long period binary stars 
in globular clusters, the binary frequency and binary period
distributions are very poorly determined. This has been
the major source of uncertainty in the theoretical modeling the role
of binaries in the evolution of
globular clusters. 
For binaries with periods in the range of 10~days to 10~years, the direct
way to determine the binary frequency is to systematically 
measure radial velocity variations in individual stars
of a globular cluster. Technically, only recently with the advent of 
high resolution multi-object spectrographs on 4-meter class 
telescopes did
it become possible for the first time to sample a large number of stars
and to obtain multi-epoch observations with a reasonable amount of telescope
time. With telescopes smaller than the Keck 10-meter telescope, we are 
limited to bright red giants and subgiants even in
the closest globular cluster. These bright low metallicity giant stars 
have smaller masses and larger radii compared to the field G-type stars.
Consequently, a binary system containing such a bright giant star is difficult
to detect since the system tends to have
a long period and small radial velocity variations.

All of the existing radial velocity surveys of binary stars in globular
clusters generally were done among the bright giant stars
(Gunn \& Griffin 1979; Pryor, Latham \&\ Hazen 1988; C$\rm \hat o$te 
{\it et al.} 1994).
The first radial velocity survey with high precision was done by 
Gunn \& Griffin (1979) in the globular cluster M~3.
They obtained a total of 85 velocity measurements for 33 
giant stars with
V magnitudes 12$-$14. 
They failed to find any 
binary candidates
among the non-pulsating stars and concluded that binaries with separations 
between 0.3$-$10~AU (periods between 0.1~years and 30~years) were much rarer 
in globular clusters than 
in the field Population I stars.
In 1988, combining Gunn \&\ Griffin's data in M~3 with new MMT
observations, 
Pryor et al. produced a larger
dataset containing 111 giant stars, each with 3 multiple measurements.  
They found one binary candidate, which later was
confirmed as a system 
with a period of 7.3~years. Recently, 
C$\rm \hat o$te et al. (1994) published a similar survey  in the 
globular cluster NGC~3201.
A total of 786 velocities were obtained for 276 giant stars with 
V magnitudes of 11.0$-$16.5 over a total timespan of 6 years.
They found 2 good binary candidates, plus 13 possible
candidates. The derived binary frequency from both surveys 
is roughly 5\% $-$ 18\% for systems with 
0.1~yrs $\le P \le$ 10~yrs
and 0.1 $\le q \le$ 1.0, depending on orbital eccentricities. 

With the advent of the Norris Multi-fiber Spectrograph
on the 200 inch telescope at the Palomar observatory, we
carried out a radial velocity survey for spectroscopic binary stars 
in the globular cluster
NGC~5053. Our survey was designed to 
reach much fainter stars than previous surveys. It is thus
sensitive to binary systems with much shorter periods, and much 
larger radial velocity variations. 
The object sample in our survey consists of 77 cluster member giant 
and subgiant stars in the 
globular cluster NGC~5053 with visual magnitudes in the 
range of 14.2$-$18.6. We have obtained 6 epochs of observations of these
77 stars in a total
timespan of 3 years. Compared to the previous radial velocity survey 
made in the same cluster by Pryor et al (Pryor, Schommer \&\ Olszewski 1991), 
our survey samples more than twice as many stars and goes almost three magnitude
fainter. Our survey is sensitive to binary systems with 
periods as short as 3 days.
For potential binaries in our sample, a maximum radial velocity 
variation as large as 
116~$\hbox{kms}^{-1}$ is possible for
an edge-on circular orbit with $M_1$ and
$M_2$ of 0.8$M_\odot$ and 0.5$M_\odot$ respectively. Our velocity
measurement error varies
from star to star, but typically is
about 3~$\hbox {km s}^{-1}$, which is somewhat worse than in previous
surveys (We discuss in detail the
velocity measurement errors and related problems in \S2.2). However, the
smaller radii of the stars
in our sample compensates for this disadvantage. In addition, our sample
stars are less luminous and don't have radial velocity variations
on a scale of 8~$\hbox{kms}^{-1}$ caused by intrinsic atmospheric
motions found among bright giant stars by previous surveys
(Gunn \&\ Griffin 1979; Lupton, Gunn \&\
Griffin 1987; Pryor {\it et al.} 1988).

The globular cluster NGC~5053 is an ultra-low density cluster with a
half-mass relaxation timescale of $\sim$ 8~Gyr (Djorgovski 1992), comparable
to the Hubble timescale.
This suggests that this cluster is barely dynamically relaxed, 
and the spatial and period distribution of its
primordial binary population
has not been significantly altered by two-body relaxation processes. 
Moreover, the long
relaxation timescale indicates that in this cluster
the binary destruction by star-binary and binary-binary encounters is 
not important. However, one disadvantage imposed on our survey is that
NGC~5053 is a very metal poor cluster with
[Fe/H] of $-$2.2, and thus the spectra from its stars have 
weaker metal absorption lines than metal rich objects. 

\bigskip
\centerline{\bf 2. THE SURVEY}
\medskip

\noindent{\bf 2.1. Observations}
\smallskip
All of the observations were made with the Norris Multi-fiber
Spectrograph on the Hale 5.0m telescope at the Palomar observatory.
The Norris Spectrograph is a fiber-fed multi-object spectrograph which is 
mounted at the Cassagrin focal plan 
of the telescope (Hamilton {\it et al.} 1993). It has a total of 176 fibers, and each fiber
has a 1.6 arcsecond diameter aperture. The small diameter
of the fiber aperture implies that the effective throughput of
the spectrograph is very sensitive to seeing conditions, especially
for stellar objects, as in our case. The fibers are located
in two opposing banks of equal number. The minimum separation
between two fibers is $\sim$ 16 arcsecond, which is the major
limitation for sampling more stars in globular clusters.
The spectrograph covers a field of view of 20$\times$20 square arcminutes.

We obtained a total of 6 epochs of observations of a sample of 77 cluster 
member stars with a total
timespan of 3 years.
The summary of all observations is tabulated in Table 1.
The first four epochs of observations in 1992 and 1993
were taken with a 1024$\times$1024 pixels CCD and the field of view 
of 10$\times$10 
square arcminutes.
The last two epochs of observations in 1994 were taken with a more sensitive 
and larger CCD with 
2048x2048~pixels.
These observations therefore had wider wavelength coverage and sampled more 
stars. Thus some of the stars in our sample have more velocity measurements 
than others. Also notice that the total exposure time for the primary field 
in NGC~5053 varies from one
epoch to another due to changes in the amount of available observing time 
and the observing conditions.
For some epochs, we could not obtain any observations for the second 
field in the same cluster. This
is one of the causes for the variations of velocity measurement errors for 
the same star at different epochs.

During all observations, we used a 1200~groove/mm grating centered around
5000~\AA. The resulting spectral scale is $\sim$ 0.65~\AA/pixel, corresponding
to the velocity scale of $\sim$39~$\hbox {km s}^{-1}\rm pixel^{-1}$. 
The spectral 
resolution of 
the observations is around 2.5$-$3~\AA.
With a cross-correlation
technique, we should be able to measure any velocity shift 
larger than one tenth of a pixel,
{\it i.e.} 4~$\hbox {km s}^{-1}$.
The poor resolution in the April 1994 observing run was due to bad collimator 
focus caused by
a mechanical problem in the spectrograph.

To make accurate velocity measurements, in all our observations we took comparison Thorium-Argon
spectra both before and after each object exposure to calibrate out any spectrograph flexure.
Dome flats were taken immediately after each setup to flatfield the object spectra. We also obtained 
some exposures of the twilight sky
to check the spectrograph velocity zero-point shifts between different 
nights and also between
different observing runs. Depending on the observing conditions and the amount 
of time we had during each run,
we observed at least the primary field and sometimes the secondary
field in NGC~5053. Since it is relative velocity variations which we 
need to measure accurately
in the search for spectroscopic binaries, we chose about 10 giant stars brighter
than 13~magnitude in the globular cluster M~13 to serve as the velocity 
standards. These giant stars 
were chosen from the sample in the radial velocity survey by Lupton, Gunn \&\ Griffin (1987). The
typical velocity error in their survey is $\sim$1~$\hbox {km s}^{-1}$. One advantage of using these stars
is that we were able to obtain all their spectra with a single exposure.

\bigskip
\noindent{\bf 2.2. Data Reduction}
\medskip
Our data were reduced using IRAF.\footnote{$^1$}{IRAF is distributed by
the National Optical Astronomy Observatories, which are operated by
the Association of Universities for Research in Astronomy, Inc., under
contract to the National Science Foundation.}
After the images were trimmed and corrected for bias, all of the object spectra,
including sky expsoures, were identified, traced and extracted using APALL 
task in the IRAF 
package SPECRED.
An optimal extraction algorithm was used to produce one-dimensional spectra
for all stars.
The dome-flat spectra from all fibers were extracted and averaged. 
A low-order polynomial
was used to fit the averaged dome-flat spectrum. The flatfield spectra were
obtained by normalizing all dome-flat spectra with the single fitted spectrum.
The CCD sensitivity
variations at small scales in all object spectra were taken out by 
dividing them with the corresponding
flatfield spectra.

Figures 1(a) and 1(b) show
spectra of the same star with a signal-to-noise ratio of $\sim$15, 
taken in 1992 
and 1994 respectively. Signal-to-noise ratios of our spectra vary
within a large range, but the typical value of a continuum is around 10. 
The spectra taken in 1994 have the wavelength coverage of 4700\AA $-$5800\AA, 
and consequently include
very strong sky emission lines such as [OI]~5577\AA, HgI 5461\AA, the Hg 
doublet 
5791\AA, and the NaD doublet 5893\AA. To obtain good
sky subtraction, we corrected the variations of throughput from 
fiber to fiber.
The correction for each fiber was estimated by dividing the integrated flux in 
the sky emission line HgI5461\AA\ with the average flux of the same 
line from all fibers.
The most useful absorption lines for doing 
radial velocity cross-correlation, H$_\beta$4861\AA\ 
and the MgI triplet at 5167, 5173 and 5184\AA, are outside
the strong sky emission line region. Residuals of sky emission 
lines in the sky-subtracted spectra
do not have a big effect on the velocity measurement.
In the cases of large residuals from the sky subtraction, we simply
mask off the regions with sky lines. Throughput correction was not 
necessary for spectra taken in 1992
since there are not many strong sky emission lines in the range of 
4710\AA$-$5300\AA. Figure 1(c) 
shows the spectrum of the radial velocity standard II-76 in the 
globular cluster M~13.

All wavelength calibrations 
were done only with a set of isolated lines. 
Typically, we used 3rd order polynomial fitting and obtained 
an rms error in the
dispersion solution less than 0.04~\AA, which corresponds to a velocity 
error of less than 2~kms$^{-1}$.
Each stellar spectrum was wavelength calibrated with its own comparison 
spectra which were taken 
through the same fiber before and after the star exposure.
To make sure spectra taken through different fibers and in different nights 
have 
the same wavelength scale, we cross-correlated every wavelength calibrated 
thorium-argon spectrum 
against all the other Th-Ar spectra. 
With care in the calibration procedure, we were able to keep the 
relative shift between all
Thorium-Argon spectra under 2~kms$^{-1}$.

The object spectra were dispersion
corrected and binned into log wavelength. Radial velocities for
all stars in the sample were obtained with a cross-correlation technique 
(Tonry \&\ Davis 1979).
Each object spectrum was cross-correlated with the template spectra of 
radial velocity
standards. We used the task FXCOR in RV package of IRAF to obtain 
radial velocity measurements. 

\bigskip
\noindent{\bf 2.3. Radial Velocities}
\medskip
We have obtained a total of 247 new radial velocities for 77 cluster member 
giant and subgiant stars in the 
globular cluster NGC~5053. Of the 77 cluster member stars, 
66 stars have multiple 
radial velocity measurements
and are suitable for a binary search. Therefore, we will concentrate 
only on those 66 stars. 
Table 2 lists photometry, astrometry and 247 radial velocities
for those 77 stars in NGC~5053. The columns record, from left to right, 
the star's
identification names, right ascension and declination in the epoch of 1950.0, 
the radial distance
from the cluster center in arcseconds, the heliocentric Julian date
(+244000 in days)
of the observation, the heliocentric radial velocity at the date and its 
corresponding uncertainty
in kms$^{-1}$, the number of observations, the weighted mean radial 
velocity and the external
velocity error in a single measurement estimated from the dispersion about 
the mean,
the chi-square for the observations and the probability of obtaining a 
chi-square at least
this large purely due to measurement errors. The final two 
columns give the star's magnitude
and (B $-$ V) color. Some of the photometry listed in Table 2 
are from Sandage
et al (Sandage, Katem \&\ Johnson 1977) and the rest are our measurements.
To distinguish the stars originally identified in the SKJ paper
from the new stars we selected, we combined the 
same names SKJ used with 
the prefix S. Table 2 is published in the CD-ROM supplement to this
journal due to its large size. Anybody who is interested in Table 2
could also contact LY directly.

In Table 2, the second and any subsequent lines under each object 
report the radial 
velocities and
the internal errors at the corresponding heliocentric Julian dates. 
For a combined set of velocities, we calculated $\chi^2$ and the
weighted mean velocity (for details, see Duquennoy \& Mayor 1991). 
The $\chi^2$ value represents the radial velocity variability over
the timespan of three years.
To evaluate the significance of the variations represented by 
$\chi^2$, 
we calculated
the probability of having a $\chi^2$ at least this large purely due to 
chance fluctuations with
a Gamma function Q(0.5$\nu$, 0.5$\chi^2$), here $\nu$ is the number of the 
degree of freedom.
Reasonable limits of $\chi^2$ probability for identifying significant 
variations are 0.01$-$ 0.001.

The velocities tabulated in Table 2 are the radial velocities of stars
in NGC5053 relative to
star II-76 in M~13.
The globular clusters M~13 and NGC~5053 have cluster systematic
velocities of about
$-$246.4~$\hbox {km s}^{-1}$
and 43~kms$^{-1}$ respectively (Pryor \&\ Meylan 1993). Figure 2 shows
the histogram of
all relative velocities tabulated in Table 2 including non-cluster
member velocities.
With only cluster members and excluding the variable stars,
the mean relative velocity is 291.0~kms$^{-1}$
and the standard deviation is
3.7~kms$^{-1}$.

\bigskip
\noindent{\bf 2.4. Error Analyses}
\medskip
As shown in Table 1, we usually took multiple exposures of the same
set of stars in a single observing run. We measured velocities using
individual spetcra as well as using the sum of these individual spetcra
of the same star. For some faint stars, summation of all exposures
within a single observing run is required in order to have enough signal
to measure velocities. 

One problem with summation is that it works fine for 
almost all types of 
potential binaries in our sample {\it except} binaries with periods of a few 
days. 
The shortest period for a binary allowed by the 
sizes of stars in our sample 
is $\sim$3~days.
In the case of an edge-on circular orbit with $M_1=0.8M_\odot, M_2=0.5M_\odot$, 
the expected peak-to-peak
radial velocity variation is $\sim$113~$\hbox {km s}^{-1}$. Obviously the 
binaries with such an orbital 
configuration
can be easily identified by inspecting the velocities
measured during the same observing run. 
We found only one binary candidate ST
with radial velocity (individual measurements) variations larger than 
30~$\hbox {km s }^{-1}$ over two days. This star is a strong candidate for
a short-period spectroscopic binary (see \S3.1 for detailed discussion of
this system).
However, the binaries with periods of a few days and {\it near 
face-on} orbits will not 
show large radial velocity variations during a single observing run. Summation
of spectra will smear out
velocity variations of this type of binaries and we can not 
distinguish them from single stars. The problem is partly inherent for 
this type of binaries,
and partly due to the noisy spectra of the faint stars in our survey.
Fortunately, the probability of having this particular type of binaries in 
our sample is very small ($\le$ $\sim$2\%). 
The final result of the survey should not be significantly affected by this 
limitation.

Making a realistic estimate of the internal error for each radial
velocity
measurement is crucial for identifying the
radial velocity variables. The internal velocity errors are evaluated as
follows. 
The quality of a 
cross-correlation can be characterized by a quantity R
(for the definition of R, see Tonry \&\ Davis 1979).
The larger
R is, the more accurate the velocity is.
The value of R reflects the signal-to-noise ratio of the object
spectrum as well as how well the 
object spectrum matches with the template spectrum.
With this definition of R, the internal error is calculated as
     $ \sigma = {\sigma_0 \over 1+R }, $
here $\sigma_0$ is a constant which can be 
estimate with our data.

The method we used to estimate $\sigma_0$ is similar to the one
described in 
Pryor et al. (1988) and Vogt et al. (1995). 
We took the velocity difference $\Delta v$
between each pair of velocities for those stars with multiple measurements. 
Then 
$\sigma_0 = {\Delta v \over [ (1+R_1)^{-2} + (1+R_2)^{-2}]^{1/2}}, $
here $R_1$ and $R_2$ are the Tonry \&\ Davis values for velocity $v_1$ 
and $v_2$ respectively. If
velocity errors have a Gaussian distribution, $\sigma_0$ is equal to the 
standard deviation
of a normal distribution produced by the above equation.
The second way is to compute the velocity difference $\Delta v$ between
the weighted mean velocity and a given measurement $v_i$.
Then $\sigma_0$ is
${(v_i - \bar v) \over [ (1+R_i)^{-2} + 
(\sum (1+R_i)^{-2} ]^{1/2}}.$
Figure 3 shows the $\sigma_0$ distribution.
It is roughly gaussian
and the standard deviation $\sigma_0$ is 44.7~km$s^{-1}$ for 
the April 1992 observation.
For all observed objects, Figure 4 is the plot of the internal 
velocity error
versus R of each velocity measurement. 
A separate $\sigma_0$ was estimated for each 
observing run. With the derived $\sigma_0$,
we got a total $\chi^2$ of 185.1 for
196 degrees of freedom for all velocities in our sample excluding non-members and potential
variables. The probability of a $\chi^2$ larger than this is 0.68, 
which is acceptable.

For all stars in our sample, velocity errors listed in Table 2 
are calculated using $\sigma_0$ estimated with the above method and
R values from the summed spetcra. 
A typical error is $\sim$3~$\hbox {km s}^{-1}$, but the worst is as 
large as 10~$\hbox {km s}^{-1}$.
This large variation is partly due to the instrinsic luminosity 
differences between stars 
in our sample, and partly
because some stars were observed more frequently than others. In addition, 
the accuracy of fiber positioning
is different from star to star and from one observing run to another. 
With non-uniform velocity measurement errors, it is dangerous to use only 
the velocity variations 
between different epochs to represent the 
true velocity variabilities due to orbital motion of binary stars. 
The more effective way of selecting binary candidates 
is to calculate $\chi^2$
of all velocity measurements and the probability of obtaining a $\chi^2$ 
value larger than observed by chance.

The velocity zero-point \ \ shifts between different epochs were examined
by cross-correlating
spectra of bright stars in M~13. The twilight spectra taken in several 
epochs 
were also used to estimate the velocity zero-point shifts. We didn't find any
significant velocity zero-point shifts between different epochs.

\bigskip
\centerline{\bf 3. RESULTS}
\bigskip

\noindent{\bf 3.1. Spectroscopic Binary Candidates}
\medskip
Using the criterion that a binary candidate must have 
a velocity $\chi^2$ such that the probability of obtaining 
a $\chi^2$ larger than this value 
by chance is less than 0.01, we identified 6 spectroscopic binary 
candidates among a sample of 66 cluster member stars. The $\chi^2$ probabilities
of these six binary candidates indicate that their radial velocity variations
are significant over the timespan of 3 years.
Of these 6 binary candidates, star S5 was previously found
as a binary candidate in the Pryor et al. survey (1991) and star ST is a
binary candidate with a very short period and a large-amplitude velocity
variation. Table 3 lists
these six spectroscopic binary candidates. 
Figure 5 is a cluster color-magnitude diagram, in which the solid triangles 
represent five of the six binary candidates
discovered by our survey, candidate ST is marked separately with a solid
square and the hollow circles show the three brighter
binary candidates
discovered in the Pryor et al. (1991) survey. 
Figures 6(a)$-$(e) are the detailed finding charts of the 6 binary candidates.
The star is 318.1$''$ away from the cluster center and located in an
isolated environment, as shown in Figure 6(e).
(B-V) color and V magnitude of binary candidate ST are
are fairly consistent with
those of the cluster member stars. 

We fit
the velocity data and obtain the orbital solutions for star ST. However, due
to the small number of velocity measurements, the orbital solution for
star ST is not unique. Figure 7 shows the three possible solutions.
In the figure, P is the period, V$_0$ the center-of-mass velocity,
K the amplitude of the orbital velocity and e the eccentricity.
These possible orbital solutions and its photometric color 
suggest that star ST is a cluster member. However, the spectrum of this
star has a much stronger Mg$\lambda$5177 triplet absorption line 
compared to that of a low-metallicity giant star. The analyses of 
the Mg$\lambda$5177 line strength verses
the photometric color (for the details of the method, see 
Faber {\it et al.} 1985; Gorgas {\it et al.} 1993) 
indicate that the spetcrum of star ST is
similar to that of a metal-rich dwarf. But considering that star ST
is a very short-period spectroscopic binary, we conclude the parcularity
of its absorption line strength could result from the possible
interaction between the two components of the binary system; star
ST is probably a cluster member. We should point out that if star ST
is a cluster member giant and if its period is around three to five
days, this binary system must be on the verge of Roche lobe overflow
since the primary star has a fairly large radius. 

\bigskip
\noindent{\bf 3.2. Modeling the binary frequency}\footnote{$^2$}{Here the binary frequency is defined
as the ratio of the number of binary candidates to the total number of 
``stars'', here ``stars'' includes
both single stars and binary stars; and a binary is counted as ``one star''.}
\medskip
In any spectroscopic survey for binary stars, the probability of detecting 
a binary depends not only on the binary frequency, but also on a set of unknown binary orbital parameters
such as period, eccentricity, mass ratio, orbital phase and inclination angle 
of the binary system. Unfavorable binary orbital configurations 
would cause a fraction 
of binary systems to be missing from our detection sample. Thus, to properly 
interpret the measurements in our 
survey, we use Monte-Carlo simulation methods to generate a large number of 
simulated radial velocity data sets and compare them with the 
measured radial velocities.
We calculate from the synthetic data sets the fraction of binaries 
missed in our survey,
then correct the 
observed binary frequency.
We applied the same
criterion used in the survey to the synthetic data for identifying 
binary candidates. This criterion
is that the probability of obtaining a $\chi^2$ larger than observed by chance
has to be less than 0.01.

An additional statistical method is to use a Kolmogorov-Smirnov (K-S) test 
to compare the 
cumulative distributions
of maximum velocity differences for the real data to the mean 
distribution determined from a 
large number of simulated data sets. This method 
has been used in several surveys, such as 
Harris \&\ McClure (1983), Pryor et al. (1988) and C$\rm \hat o$te 
et al. (1994). We will 
discuss in detail the applicability of this method to our data 
in section \S4.3.

\medskip
\noindent{\bf 3.2.1. The binary models}
\smallskip
The simulation code for generating synthetic radial velocities
was generously provided to us by Dr. C. Pryor. Some modifications have been
made during our calculations. In solving Kepler's equations, we
made the following assumptions: the binary period, mass ratio and 
eccentricity distribution functions were taken as the ones in 
Duquennoy and Mayor (1991, thereafter DM), which were derived for
G-type dwarf stars in the solar neighbourhood. The orbital 
longitude $\omega$ and the initial orbital phase were drawn
randomly between 0 and 2$\pi$ from a uniform distribution. 
$\cos(i)$, where i is the inclination angle of the orbital plan to the
line of sight, was chosen randomly from 0 to 1 with a uniform
distribution. Of course, the mass of the primary is assumed to be 
0.8$\rm M_\odot$. 
Our survey is sensitive to binaries with the minimum and maximum periods 
of 3~days and 10~years respectively, and with mass ratio larger than
0.125.

The effect of mass transfer was considered in our simulations
by eliminating the binary system whenever the two stars get closer than
the critical Roche lobe radius. The radius of the primary was calculated
from its photometry using the Revised Yale Isochrones (Green, 
Demarque \&\ King 1987). Figure 8 shows the stellar radius versus
absolute visual magnitude (reddening corrected) for the stars in our
sample. In Figure 8, the solid line is calculated for the metallicity of
$-2.2$ from the models, the solid triangles represent the stars in our 
sample with the adopted distance modulus $\rm (m - M)_V$$_0$ of
16.08~magnitude and $\rm E(B-V)$ of 0.06~magnitude (Fahlman, Richer \&\
Nemec 1991). Also shown in the figure with the hollow squares are the
empirical radii for giants in the clusters M~92 by Cohen, Frogel and
Persson (1987) based on broad-band infrared photometry and narrow-band
CO and H$_2$O indices. The globular cluster M~92 has a metallicity
[Fe/H] of $-$2.2, very similar to NGC~5053. 

We didn't consider the effect of asymptotic giant branch in our
calculations because in our sample less than 24\% of stars are above the
Horizontal branch. 

\medskip
\noindent{\bf 3.2.2. The binary frequency}
\medskip
To convert the binary discovery fraction of
0.084 (5.5/65.5, giving candidate ST 0.5 weight due to some ambiguity of
its cluster membership) to the true binary frequency $f_b$, we estimate
the incompleteness correction using Monte-Carlo simulations.
This correction is also called the binary discovery 
efficiency $D_b$, which is defined as the fraction 
of the ``discovered'' binaries in the synthetic velocity data if we assume 
all 66 stars as binary systems.
Radial velocities
of each binary were calculated using the binary models described in \S3.2.1 
at the actual observation 
dates. Velocity noises were drawn randomly from Gaussian distributions  
with the mean of zero and standard deviations of the {\it real velocity 
measurement errors}.
By applying the same binary identifying criterion used in the survey, 
we obtained the fraction of ``discovered'' binaries in the simulated data.
Figures 9(a)$-$(b) show the binary discovery efficiency $D_b$ as a function of
period $P$ in days for the binary models with circular orbits and with 
eccentricity distribution $f(e) = 1.5\sqrt e$ respectively.
Notice that each data point in the plots is the mean 
of 1000 simulation tries.
Here we assume the primary star in a binary system to have 0.8M$_\odot$,
and a mass ratio q (defined as M$_2$/M$_1$) 
in the range of 0.125 to 1.75. This implies that the 
secondary could be anything from
a 0.1M$_\odot$ low mass star to a 1.4M$_\odot$ heavy neutron star. 
In both plots,
the solid triangles and dots represent simulations with and without the 
effect of mass transfer
respectively. It is clearly shown in the figures that the biasing of the 
discovered binary orbits  by the size
of the giants is significant at the short period end. This bias is 
higher for stars with brighter 
magnitudes,
which is the major reason that the previous surveys could not detect any binary with period shorter
than 40~days.

In the case of non-zero eccentricity orbits, the average binary discovery 
efficiency is about 29\%\ for binaries with 
3~days $\le P \le$ 10~years and 0.125 $\le q \le$ 1.75. For binaries with
the same period and mass ratio range, the inferred true  
binary frequency $f_b$ is 0.084/D$_b$, i.e. 29\%.
In the case of circular orbits, the corresponding binary frequency
for systems with periods and mass ratios in the same ranges is 26\%.
The discovery efficiency varies as the limits of binary period and mass ratio 
change. A binary with a very low mass companion is usually difficult to 
detect even if the period is short. 
For instance, the primary star in a binary with a period 
of 100~days shows a maximum
radial velocity change of 10~$\hbox {km s}^{-1}$ in the case of 
an edge-on circular orbit and 
$M_1 = 0.8M_\odot, M_2=0.1M_\odot$. 
If we choose the sensitivity limits of our survey as 3~days $\le P \le 10$~years and $0.3 \le q \le 1.75$, 
the binary discovery 
efficiency is 35\%\ and 37\%\ for the binary models with non-circular 
and circular orbits
respectively. The corresponding binary frequency $f_b$ in the cluster is 24\%\ 
and 23\%\ respectively.

\bigskip
\centerline{\bf 4. DISCUSSION}
\medskip
\noindent{\bf 4.1. An alternative period distribution function in globular clusters}
\medskip
The binary frequency estimated above depends on the adopted period distribution
in the simulations.
An alternative period distribution
has a constant number of binaries per unit logarithmic period interval.
This so called ``flat'' period distribution is a crude representation of
the survey by Abt \&\ Levy (1976). 
The ``flat'' distribution produces more short-period binaries and less 
long period
ones for 3~d $\le P \le $10~yr than the DM period distribution. Obviously, it 
also gives a somewhat higher binary discovery efficiency for our survey. 
Specifically, with an average discovery efficiency of 36\% derived 
from the ``flat''
distribution, we obtained a binary frequency of 23\% for systems with 3~d $\le P \le $10~yr,
$0.125 \le q \le 1.75$ and $f(e) = 1.5\sqrt e$. Similarly in the case of
circular orbits, the inferred 
binary frequency is 21\%. For the mass ratio range of $0.3 \le q \le
1.75$, the derived binary frequencies are 19\%\ and 22\%\ 
for circular and eccentric orbits respectively.
The binary frequency derived by using the ''flat'' period distribution
is smaller
than using the DM period distribution in the simulations. The 
binary period distribution in globular clusters perhaps bears more
similarities with ones in open clusters and among Pop. II low 
metallicity halo stars. Many extensive programs of studying 
binary stars in open clusters and among low metallicity halo stars 
(Carney \&\ Latham 1987; Latham {\it et al.} 1988, 1992) should 
shed light on the properties of binary populations in globular
clusters.

\medskip
\noindent{\bf 4.2. The effect of stellar encounters}
\medskip
We did not take into account the effect of stellar encounters 
in our calculations.
This effect can alter the shape of the primordial distributions of 
binary orbital elements in globular clusters. 
As briefly discussed in \S2.1, 
this effect is not significant
in very low density clusters such as NGC~5053 since the cluster half-mass 
relaxation timescale is comparable to
a Hubble timescale $T_h$.
However, of the eight globular clusters in which radial velocity surveys 
for binary stars have been carried out 
by various groups (Hut {\it et al.} 1992), 
six have cluster central
relaxation timescales $T_{rh}$
much shorter than $T_h$ of 14~Gyr.
These six clusters include
M~3, M~13, 47~Tuc, M~2, M~71 and NGC~3201, where $T_{rh}$ is in the 
range of $10^8$ to 
$10^9$~yrs. NGC~3201 and M~71 have particularly short $T_{rh}$ of 
only 100 million years
in spite of their apparent low central densities. 

It is a rather complicated problem to characterize 
quantitatively the change
in the primordial period distribution due to dynamical evolution.
A tremendous amount of theoretical computations have been devoted 
to this subject (Phinney \&\ Sigurdsson 1991;
McMillan \&\ Hut 1994; Sigurdsson \&\ Phinney 1995). 
The detailed N-body simulations (McMillan, Hut \&\
Makino 1991) show that binary-binary encounters are a very effective 
binary-destruction process. 
Most binary destruction occurs within a few core radii of the cluster center.
In addition, some binaries can be ejected to the outer parts of a cluster
due to large recoil during binary-binary encounters in the core.
Thus, in dynamically evolved clusters such as M~71 it
is perilous to assume the field
period distribution.

Although it is premature to draw any comprehensive conclusions about 
the binary 
frequency dependence on the dynamical properties of globular clusters, 
it is illuminating
to compare the results from various surveys for binary stars in different
globular clusters. As first pointed out by Pryor et al (1991; Hut {\it
et al.} 1992), the binary discovery rate in NGC~5053 is notably higher
than other clusters where binary searches have been carried out.
The result of our survey appears to further support Pryor's 
conclusion. This is perhaps associated with the fact that NGC~5053 is
a dynamically young cluster. 

\medskip
\noindent{\bf 4.3. Kolmogorov-Smirnov (K-S) tests}
\medskip
As Harris \&\ McClure (1983) first pointed out, small number statistics
are important in estimating binary frequencies in globular clusters. 
Subsequently a sophisticated 
and quantitative statistical method ---the Kolmogorov-Smirnov test---
has been employed in many surveys.
K-S tests were applied to the cumulative
distributions of maximum velocity difference obtained from both simulated 
and real data. A K-S
statistic is useful for {\it rejecting} the null hypothesis
that two data sets are drawn from the same parent distribution. 

We applied K-S statistics to our measured velocities 
and the simulated data. 
In Table 4, we tabulated the confidence level
at which the null 
hypothesis of the simulated data and real data drawn from the same 
distribution is accepted. K-S tests show that the binary frequency in NGC~5053
is close to 25\%, and the hypothesis that the binary frequency
is higher than 50\%\ is rejected with confidence higher than 
85\%. However, it is noticed in Table 4 that with 
our data K-S tests are not very effective
in {\it rejecting} the cases with different binary frequencies. 
This problem results from some of the large errors in our 
velocity measurements. 
As described previously, the synthetic data were generated using the actual
velocity errors. The corresponding cumulative distribution of maximum
velocity variation is primarily controlled by a few large errors.
K-S tests have been applied more effectively by Pryor et al. (1989)
and C$\rm \hat o$te et al. (1994) to their data with the radial velocity 
errors are as small
as 1~kms$^{-1}$. They have derived a binary frequency in the
range of 10\% to 20\% for binaries with 0.1~yr $\le$ P $\le$ 20~yr.

We would like to thank The Kenneth T. and Eileen L. Norris Foundation
and especially Kenneth Norris for their generous support in building
the Norris spectrograph. Michael Doyle, John Henning and Juan Carrasco
of the Palomar Observatory staff are thanked for providing excellent 
service during 
the observations. We wish to thank Tad Pryor for his unreserved
help and interesting discussions about astronomy
on many occasions. The Norris Spectrograph was built by Bev Oke, Judith 
Cohen and Donald Hamilton; we thank Donald Hamilton and Todd Small 
for their help. We also wish to thank an anonymous referee for providing 
many excellent suggestions
which have helped to improve
the paper. 

\vfill\eject
\bigskip
\noindent{\bf REFERENCES}
\bigskip
\refindent Abt, H.A. \&\ Levy, S.G., 1976, ApJS, 30, 273
\refindent Carney, B.W. \&\ Latham, D.W., 1987, AJ, 93, 116
\refindent Cohen, J.G., Frogel, J.A., \&\ Persson, S.E., 1978, ApJ, 222, 165
\refindent C$\rm \hat o$te, P., Welch, D.L., Fischer, P., 
Da Costa, G.S., Tamblyn, P., Seitzer, P., \&\ Irwin, M.J., 1994,
ApJS, 90, 83
\refindent Djorgovski, S. \&\ King, I.R., 1986, ApJL, 305, L61
\refindent Djorgovski, S., 1992, 
in {\it Structure and Dynamics of Globular Clusters}, ed. by
S.G. Djorgovski and G. Meylan, (San Fransisco: ASP), 50, 357
\refindent Duquennoy, A. \&\ Mayor, M., 1991, A\&A, 248, 485 
\refindent Faber, S.M., Friel, E.D., Burstein, D. \&\ Gaskell, C.M., 1985, ApJS, 57, 711
\refindent Fahlman, G.G., Richer, H.B. \&\ Nemec, J., 1991, ApJ, 380, 124
\refindent Gao, B., Goodman, J., Cohn, H., Murphy, B.,
1991, ApJ 370, 567
\refindent Gorgas, J., Faber, S.M., Burstein, D., Gonzalez, J.J., Courteau, S. \&\ Prosser, C., 1993, ApJS, 86, 153
\refindent Green, E.M., Demarque, P. \&\ King, C.R., 1987, {\it The Revised Yale Isochrones and Luminosity
Functions}, Yale University Observatory.
\refindent Grindlay, J.E., Hertz, P., Steiner, J.E., Murray, S.S. \&\
Lightman, A.P., 1984, ApJ, 282, L13
\refindent Gunn, J.E. \& Griffin, R.F., 1979, AJ, 84, 752
\refindent Kaluzny, J. \&\ Krzeminski, W., 1993, MNRAS, 264, 785
\refindent Kulkarni, S.R., Anderson, S.B., Prince, T.A. \&\ Wolszczan, A., 1991,
 Nature, 349, 47
\refindent Hamilton, D., Oke, J.B., Carr, M.A., Cromer, J., Harris, F.H., Cohen, J.G., Emery, E. \&\
Blakee, L., 1993, PASP, 105, 1308
\refindent Harris, H.C. \&\ McClure, R.D., 1983, ApJL, 265, L77
\refindent Heggie, D.C. \&\ Aarseth, S.J., 1992, MNRAS, 257, 513
\refindent Hut, P., McMillan, S.L.W., Goodman, J., Mateo, M., Phinney,
E.S., Richer, H.B., Verbunt, F. and Weinberg, M., 1992, PASP, 104, 981
\refindent Latham, D.W.,\ \ Mazeh, T., Stefanik, R.P., Davis, R.J., Carney, 
B.W., Krymolowski, Y., Laird, J.B., Torres, G. \&\ Morse, J.A., 1992, AJ, 
104, 774
\refindent Latham, D.W., Mazeh, T., Carney, B.W., McCrosky, R.E., Stefanik, 
R.P. \&\ Davis, R.J., 1988, AJ, 96, 567
\refindent Lupton, R.H., Gunn, J.E. \&\ Griffin, R.F., 1987, AJ, 93, 1114
\refindent Mateo, M., Harris, H.C., Nemec, J. and Olszewski, E.W.,
1990, AJ, 100, 469.
\refindent McKenna, J. \&\ Lyne, A.G., 1988, Nature, 336, 226
\refindent McMillan, S.L.W. \&\ Hut, P., 1994, ApJ, 427, 793
\refindent McMillan, S.L.W., Hut, P. \&\ Makino, J., 1991, ApJ, 372, 111
\refindent Phinney, E.S., 1992, Philos. Trans. R. Soc. Lond., A, 341, 39
\refindent Phinney, E.S. \&\ Sigurdsson, S., 1991, Nature, 349, 220
\refindent Phinney, E.S., 1996, in {\it The Origins, Evolution, and
Destinies of Binary Stars in Clusters}, ed.
Eugene F. Milone, (San Francisco: ASP), 90, in press
\refindent Pryor, C.P., Latham, D.W. \&\ Hazen, M.L., 1988, AJ, 96, 123
\refindent Pryor, C.P., Meylan, G. 1993, in {\it Structure and Dynamics of Globular Clusters}, ed. by 
S.G. Djorgovski and G. Meylan, (San Francisco: ASP), 50, 357
\refindent Pryor, C., Schommer, R.A. \&\ Olszewski, E.W. 1991, in {\it The Formation and Evolution of 
Star Clusters}, ed. Janes, (San Fransisco: ASP), 13, 121
\refindent Sandage, A.R., Katem, B. \&\ Johnson, H.L., 1977, AJ, 82, 389
\refindent Sigurdsson, S. \&\ Phinney, E.S., 1995, ApJS, 99, 609
\refindent Tonry, J. \&\ Davis, M., 1979, AJ, 84, 1511
\refindent Vesperini, E. \&\ Chernoff, D.F., 1994, ApJ, 431, 231
\refindent Vogt, S., Mateo, M., Olszewski, E.W. \&\ Keane, M.J., 1995,
AJ, 109, 151
\refindent Yan, L. \&\ Mateo, M., 1994, AJ, 108, 1810
\refindent Yan, L. \&\ Reid, N. I., 1995, MNRAS, 279, 751

\vfill\eject
\noindent{\bf FIGURE CAPTIONS:}
\smallskip
\noindent{Figure 1: Figures 1(a) and 1(b) show spectra of star LY014
taken in 1992 and 1994.
Figure 1(c) is the spectrum of star II-76 in M~13 which serves as a
radial velocity standard.}
\medskip
\noindent{Figure 2: The
histogram of all relative radial velocities measured
in our survey including non-members of the cluster stars. Each bin is
1.55$\hbox{km s}^{-1}$ wide.
The majority of the stellar radial velocities
distributed around the cluster relative velocity.
Non-members of the cluster stars are clearly distinguished from the
member stars by their radial velocities. Also notice that the dispersion
around the mean radial velocity is only 3.7~kms$^{-1}$ excluding the
measurements from non-member
stars and binary candidates.}
\medskip
\noindent{Figure 3: The distribution of
$\sigma_0$ 
derived from our data. See the text for the detailed discussion.}
\medskip
\noindent{Figure 4: The plot shows the velocity error versus the
Tonry \&\ Davis value R. The envelope of the velocity errors in the
figure
is roughly proportional to $\propto R^{-1}$, as described in
Tonry \&\ Davis (1979).}
\medskip
\noindent{Figure 5: This shows the cluster
color-magnitude diagram, where the solid triangles show
five of the six binary
candidates discovered in this survey, the solid square indicates the
short-period
binary candidate ST, and the open circles represent the
candidates found previously by Pryor et al. (1991).}
\medskip
\noindent{Figure 6: The detailed finding charts for the
6 binary candidates. All objects are plotted relative to the position
with
RA (1950)$=$13:14:5.070 and DEC (1950)$=$17:56:38.00. The axes are in
units of arcsecond, the north and
the east are indicated in the plot.}
\medskip
\noindent{Figure 7: The plots show the three possible orbital solutions
fitted to the available data.}
\medskip
\noindent{Figure 8: This is a plot of the stellar radius vs. the
luminosity for stars
in globular clusters.
The curve is
generated using the Revised Yale
Isochrones
and Luminosity Functions (Green {\it et al.} 1987)
with Y$=0.2$ and an age of 15~Gyr. The solid traingles represent the
stars in our sample, and open squares are the measurements for stars in
M~92 by Cohen et al. (1987).}
\medskip
\noindent{Figure 9: Figure
shows the binary discovery efficiency
D$_b$ as a function of period P (in days) for two different binary
models.
In both panels the solid triangles and the solid dots represent the
simulations with and without considering
the mass transfer effect respectively. 
Note that each data
point in the plots
is the mean of 1000 simulation tries.}

\vfill\eject
\nopagenumbers
\baselineskip=15pt
\centerline{\bf TABLE 1}
\bigskip
\centerline{\bf Summary of Observing Log}
\bigskip
\settabs\+DateMMMM & ObjectMM & ExptMMMMM & waveleng & ResolutionM & Seeing & \cr
\hrule\vskip2pt\hrule\vskip5pt
\+Date & Object & \ \ \ \ Exp. Time & \ \ \ \ $\Delta \lambda $ & Resolution & Weather & \cr
\+(UT) & & \ \ \ \ \ \ \ (sec) & \ \ \ \ (\AA) & \ \ \ \ (\AA) & conditions & \cr
\vskip10pt\hrule\vskip10pt
\+7/4/1992 & NGC~5053 & \ \ \ \ \ \ 3$\times$3000 & 4710$-$5310 & \ \ \ \ \ 2.5 & clear, 1.3$''$ seeing & \cr
\+7/4/1992 & M~13 & \ \ \ \ \ \ 1$\times$900 & 4710$-$5310 & \ \ \ \ \ 2.5 & clear, 1.3$''$ seeing& \cr
\+8/4/1992 & NGC~5053$^2$ & \ \ \ \ \ \ 3$\times$3000 & 4710$-$5310 & \ \ \ \ \  2.5 & cloudy, 1.3$''$ seeing & \cr
\+8/4/1992 & M~13 & \ \ \ \ \ \  1$\times$900 & 4710$-$5310 & \ \ \ \ \ 2.5 & cloudy, 1.3$''$ seeing  & \cr
\+25/5/1992 & NGC~5053 & \ \ \ \ \ \ 2$\times$3000 & 4710$-$5310 & \ \ \ \ \ 2.5 & cloudy, 1.5$''$ seeing  & \cr
\+26/5/1992 & NGC~5053 & \ \ \ \ \ \ 3$\times$3000 & 4710$-$5310 & \ \ \ \ \ 2.5 & clear, 1.5$''$ seeing & \cr
\+26/5/1992 & M~13 & \ \ \ \ \ \ 1$\times$900 & 4710$-$5310 & \ \ \ \ \ 2.5 & clear, 1.5$''$ seeing & \cr
\+16/4/1993 & NGC~5053 & \ \ \ \ \ \ 1$\times$3000 & 4710$-$5310 & \ \ \ \ \ 2.5
 & cloudy, 1.5$''$ seeing & \cr
\+16/4/1993 & M~13 & \ \ \ \ \ \ 1$\times$900 & 4710$-$5310 & \ \ \ \ \ 2.5& cloudy, 1.5$''$ seeing & \cr
\+21/5/1993 & NGC~5053 & \ \ \ \ \ \ 1$\times$2515 & 4710$-$5310 & \ \ \ \ \ 2.5
 & cloudy, 1.5$''$ seeing & \cr
\+21/5/1993 & M~13 & \ \ \ \ \ \ 1$\times$900 & 4710$-$5310 & \ \ \ \ \ 2.5& cloudy, 1.5$''$ seeing & \cr
\+11/4/1994 & NGC~5053 & \ \ \ \ \ \ 3$\times$3000 & 4700$-$5800 & \ \ \ \ \  3.0 & clear, 1.2$''$ seeing & \cr
\+11/4/1994 & M~13 & \ \ \ \ \ \ 2$\times$700 & 4700$-$5800 & \ \ \ \ \ 3.0 & clear, 1.2$''$ seeing & \cr
\+12/4/1994 & NGC~5053 & \ \ \ \ \ \ 3$\times$3000 & 4700$-$5800 & \ \ \ \ \ 2.5 & cloudy, 1.5$''$ seeing & \cr
\+12/4/1994 & M~13 & \ \ \ \ \ \ 3$\times$900 & 4700$-$5800 & \ \ \ \ \ 3.0 & cloudy, 1.5$''$ seeing & \cr
\+12/4/1994 & NGC~5053$^2$ & \ \ \ \ \ \ 4$\times$3000 & 4700$-$5800 & \ \ \ \ \ 3.0 & cloudy, 1.5$''$ seeing & \cr
\+8/5/1994 & NGC~5053 & \ \ \ \ \ \ 3$\times$3000 & 4700$-$5800 & \ \ \ \ \ 2.5 & cloudy, 1.3$''$ seeing & \cr
\+8/5/1994 & M~13 & \ \ \ \ \ \ 1$\times$600 & 4700$-$5800 & \ \ \ \ \ 2.5 & cloudy, 1.3$''$ seeing & \cr
\+10/5/1994 & NGC~5053 & \ \ \ \ \ \ 2$\times$3000 & 4700$-$5800 & \ \ \ \ \ 2.5 & clear, 1.2$''$ seeing & \cr
\+10/5/1994 & M~13 & \ \ \ \ \ \ 3$\times$600 & 4700$-$5800 & \ \ \ \ \ 2.5& clear, 1.5$''$ seeing & \cr
\+10/5/1994 & NGC~5053$^2$ & \ \ \ \ \ \ 2$\times$3000 & 4700$-$5800 & \ \ \ \ \ 3.0 & cloudy, 1.5$''$ seeing & \cr
\+10/5/1994 & M~13 & \ \ \ \ \ \ 1$\times$900 & 4700$-$5800 & \ \ \ \ \ 2.5 & clear, 1.3$''$ seeing & \cr
\vskip5pt\hrule
\bigskip
\noindent Notes: NGC~5053$^2$ is the second field we observed in the same cluster. 
In M~13 several bright giant stars were observed to serve as 
the radial velocity standards. The radial velocities of these stars
have been accurately measured by Lupton, Gunn and Griffin (1987). 
The third column in the table gives the number of exposure times the time of a single 
exposure, {\it i.e.} the total 
exposure time.
The spectral resolution in April 1994 is larger than other runs due to problems with the 
collimator focus in the Norris spectrograph during
that observing run. The CCD spectral scale is 0.65~\AA\ pixel$^{-1}$ for all of the observations.

\vfill\eject

\centerline{\bf TABLE 4}
\bigskip
\centerline{\bf KOLMOGOROV-SMIRNOV PROBABILITIES}
\centerline{\bf FOR MODEL ACCEPTANCE}
\bigskip
\settabs\+Quantity & MMMMMMMMMcase1MMMMMMMMMMMM & \cr
\hrule\vskip2pt\hrule\vskip5pt
\+ \ \ \ \ \ \ \ \ \ \ \ \ \ \ & \ \ \ \ \ \ \ \ \ \ \ \ \ \ \ \ \ \ \ \ \ \ \ \ \ \ 3d $\le$ P $\le$ 10yr & \cr
\+ \ \ \ \ \ \ \ \ \ \ \ \ \ \ &  \ \ \ \ \ \ \ \ \ \ \ \ \ \ \ \ \ \ \ \ \ \ \ \  0.125 $\le $q $\le$ 1.75 & \cr
\+ \ \ \ \ \ \ \ \ \ \ \ \ \ \ & \ \ \ \ \ \ \ \ \ \ \ \ \hrulefill \  & \cr 
\+ \ \ \ \ \ \ \ \ \ \ \ \ \ \ $f_b$ & \ \ \ \ \ \ \ \ \ \ \ \ \ \ \ \ \ \ \ \ \ \ $e=0.0$ \ \ \ \ \   $f(e) = 1.5\sqrt e$ & \cr
\vskip10pt\hrule\vskip5pt
\+ \ \ \ \ \ \ \ \ \ \ \ \ \ \ 0\%\ & \ \ \ \  \ \ \ \ \ \ \ \ \ \ \ \ \ \ \ \ \ \ \ \  0.47 \ \ \ \ \ \ \ \ \ \ 0.48 & \cr
\+ \ \ \ \ \ \ \ \ \ \ \ \ \ \ 10\%\ & \ \ \ \ \ \ \ \ \ \ \ \ \ \ \ \ \ \ \ \ \ \ \ \ 0.69  \ \ \ \ \ \ \ \ \ \ 0.72 & \cr
\+ \ \ \ \ \ \ \ \ \ \ \ \ \ \ 20\%\ & \ \ \ \ \ \ \ \ \ \ \ \ \ \ \ \ \ \ \ \ \ \ \ \ 0.88  \ \ \ \ \ \ \ \ \ \ 0.89 &  \cr
\+ \ \ \ \ \ \ \ \ \ \ \ \ \ \ 25\%\ & \ \ \ \ \ \ \ \ \ \ \ \ \ \ \ \ \ \ \ \ \ \ \ \ 0.92  \ \ \ \ \ \ \ \ \ \ 0.92 & \cr
\+ \ \ \ \ \ \ \ \ \ \ \ \ \ \ 30\%\ & \ \ \ \ \ \ \ \ \ \ \ \ \ \ \ \ \ \ \ \ \ \ \ \ 0.81  \ \ \ \ \ \ \ \ \ \ 0.82 & \cr
\+ \ \ \ \ \ \ \ \ \ \ \ \ \ \ 40\%\ & \ \ \ \ \ \ \ \ \ \ \ \ \ \ \ \ \ \ \ \ \ \ \ \ 0.46  \ \ \ \ \ \ \ \ \ \ 0.45 & \cr
\+ \ \ \ \ \ \ \ \ \ \ \ \ \ \ 50\%\ & \ \ \ \ \ \ \ \ \ \ \ \ \ \ \ \ \ \ \ \ \ \ \ \ 0.14  \ \ \ \ \ \ \ \ \ \ 0.16 & \cr
\vskip5pt\hrule

\vfill\eject

\bye